\documentclass[superscriptaddress,nofootinbib,twocolumn]{revtex4}
\usepackage{amsmath,lscape,epsfig}
\usepackage{amsfonts}
\usepackage{amsmath}
\usepackage{amssymb}
\usepackage{slashed}
\usepackage[utf8]{inputenc}
\usepackage{graphicx}
\usepackage{epstopdf}
\usepackage{color}
\usepackage[usenames,dvipsnames]{xcolor}
\usepackage{ulem}
\usepackage{comment}
\usepackage{enumitem}

\def\beq{\begin{equation}}
\def\eeq{\end{equation}}
\def\beqa{\begin{eqnarray}}
\def\eeqa{\end{eqnarray}}
\def\ban{\begin{eqnarray*}}
\def\ean{\end{eqnarray*}}
\def\bi{\begin{itemize}}
\def\ei{\end{itemize}}

\begin{document}

 \title{Artificial first-order phase transition in a magnetized Nambu--Jona-Lasinio model with a quark anomalous magnetic moment}

\author{William R. Tavares}
\email{tavares.william@ce.uerj.br}
\affiliation{Departamento de Física Teórica, Universidade do Estado do Rio de Janeiro, 20550-013 Rio de Janeiro, RJ, Brazil}  
\affiliation{Departamento de F\'{\i}sica, Universidade Federal de Santa
  Catarina, 88040-900 Florian\'{o}polis, SC, Brazil } 

\author{Sidney S. Avancini}
\email{sidney.avancini@ufsc.br}
\affiliation{Departamento de F\'{\i}sica, Universidade Federal de Santa
  Catarina, 88040-900 Florian\'{o}polis, SC, Brazil }

\author{Ricardo L. S. Farias}
\email{ricardo.farias@ufsm.br}
\affiliation{Departamento de F\'{\i}sica, Universidade Federal de Santa Maria,
97105-900 Santa Maria, RS, Brazil}

\author{Rafael P. Cardoso}
\email{cardoso.rafael@posgrad.ufsc.br}
\affiliation{Departamento de F\'{\i}sica, Universidade Federal de Santa
  Catarina, 88040-900 Florian\'{o}polis, SC, Brazil }

 \begin{abstract} 
 Recently, first-order phase transitions have been predicted as an effect of the inclusion of quark anomalous magnetic moment (AMM) in the hot and magnetized Nambu--Jona-Lasinio model (NJL). These transitions appear in the chiral condensate for different combinations of AMM and magnetic fields and could lead to inverse magnetic catalysis. However, in this work, we show that the predicted first-order phase transitions are related to regularization-dependent issues. To show this, we explore, in the context of the vacuum magnetic regularization (VMR) scheme, two different scenarios: when mass-dependent (MD) and mass-independent (MI) terms are present in the subtraction of the divergences. In the MD case, as we increase the AMM value, it is observed the appearance of a nonmassive minimum in the thermodynamical potential, which induces a first-order phase transition from the massive minimum. We argue that the MD terms must be avoided in order to satisfy the predictions of Lattice QCD, and we propose a MI solution that is valid in the limit which the magnetic fields are smaller than the squared of vacuum effective quark mass.
 \end{abstract}

\maketitle

\section{Introduction}
In the nonpertubative regime of quantum chromodynamics (QCD), the anomalous chromomagnetic moment appears as a result of the dynamical chiral symmetry breaking for massive quarks, which consequently enables quarks to possess an anomalous magnetic moment (AMM) \cite{Chang:2010hb}. It is possible, however, in a more simple way, to estimate the quark AMM from some quark models \cite{Bicudo:1998qb,Fayazbakhsh:2014mca} through the experimental knowledge of the proton and neutron magnetic moments. This simplification has stimulated investigations of the magnetized QCD phase diagram since one can explore how it affects the chiral and deconfinement phase transitions, as well as the effect of inverse magnetic catalysis (IMC) \cite{Bali:2012zg}.

The Nambu--Jona-Lasinio (NJL) model \cite{Nambu:1961tp,Nambu:1961fr}, for its simplicity, is one of the most used quark models to describe the QCD phase diagram. It is well adaptable to the inclusion of constant external magnetic fields, that are expected to be present in peripheral heavy ion collisions \cite{Kharzeev:2007jp} and magnetars \cite{Duncan:1992hi}. As a nonrenormalizable model, the use of an unsatisfactory regularization procedure can be a source of strong nonphysical behavior in the quark condensate \cite{Avancini:2019wed} and its thermodynamic properties \cite{Avancini:2020xqe}. This is the case of using simple form-factor regularizations in the one-dimensional integrations, due to the Landau Levels quantization, which will entangle the magnetic field contributions with the vacuum \cite{Avancini:2019wed}. It is well known that the magnetic field independent regularization (MFIR) methods avoid these artificial results due to the full separation of finite magnetic field contributions from the cutoff, $\Lambda$, of the theory \cite{Avancini:2019wed,Menezes:2009uc}. It is possible, however, to work within a regularization prescription useful for situations when some entangled vacuum-magnetic terms cannot be avoided, this is the case of the vacuum magnetic regularization (VMR) prescription, which is based in the MFIR, applied recently for zero \cite{Avancini:2020xqe} and nonzero AMM environments \cite{Farias:2021fci}.

All previous issues are supposed to be present in different situations concerning the magnetized quark matter in the $3+1$ D NJL-type models. When studying the influence of the quark AMM through the Schwinger {\it ansatz} \cite{Fayazbakhsh:2014mca}, most applications in the NJL model make use of non-MFIR methods \cite{Fayazbakhsh:2014mca,Chaudhuri:2019lbw,Ghosh:2020xwp,Mei:2020jzn,Xu:2020yag,Chaudhuri:2020lga,Ghosh:2021dlo,Chaudhuri:2021skc,Chaudhuri:2021lui,Wen:2021mgm,Kawaguchi:2022dbq,Lin:2022ied,Mao:2022dqn,Chaudhuri:2022oru}, besides some attempts to obtain some similar application of the subtraction scheme of divergences \cite{Aguirre:2020tiy,Aguirre:2021ljk}. 
In the non-MFIR applications, recent results have been predicted, as the possibility of a first-order phase transitions in the chiral condensate as a function of the magnetic field, mainly when considering very strong magnetic fields or a substantial value of the quark AMM \cite{Fayazbakhsh:2014mca,Kawaguchi:2022dbq} and inverse magnetic catalysis at zero temperature \cite{Chaudhuri:2019lbw,Ghosh:2020xwp}, all of them accompanied with strong oscillations characteristic of non-MFIR methods. Although all of these results represent new possibilities, most of them are not predicted by Lattice QCD.

Our aim in this work is to apply the VMR regularization prescription to the usual mean-field thermodynamic potential of the magnetized two-flavor NJL model with a constant AMM influence. Our results are contrasted with the literature in two central points:

\begin{enumerate}[label=(\roman*)]
 \item   We identify mass-dependent (MD) divergent terms proportional to $\mathcal{O}(eB^2)$ in the Taylor expansion of the ultraviolet regime in the thermodynamic potential, $\Omega$. Then we need to subtract and regularize these mass-dependent divergent contributions that cannot be ignored, since they modify the phase diagram structure, guided by the chiral condensate, $\langle\bar{\psi}\psi\rangle\sim M$, through $\partial \Omega/\partial M=0$. The thermodynamic potential at a strong magnetic field (or strong value of quark AMM) is then unbalanced to lower values in the region of $M=0$ until it becomes lower than the massive minimum, inducing the idea of a possible first-order phase transition. We show that this effect is due to the contribution of these MD terms.
 
 \item  To solve these problems observed in the previous item we propose that the Schwinger {\it ansatz} is applicable to the situation when $(|q_feB|/M_0^2)\rightarrow 0$, with $M_0$ being the vacuum effective quark mass. The Lagrangian obtained in this limit is the one used by Dittrich in Ref. \cite{Dittrich:1977ee} recently used in \cite{Farias:2021fci} and the additional MD terms observed in the ultraviolet limit of the thermodynamic potential are eliminated.
\end{enumerate}

This work has the following structure: in Sec. \ref{formalism} we present the two-flavor NJL model with the inclusion of the AMM. In Sec. \ref{pot_eff} the analytical details to obtain the thermodynamic potential are discussed. In Sec. \ref{MDreg} the MD regularization for the thermodynamic potential is shown.
In Sec. \ref{results} we present the numerical results and in Sec. \ref{secMI} we show in detail the mass-independent (MI) regularization procedure 
and in Sec. \ref{conclusions} the conclusions.

\section{Two-flavor Nambu--Jona-Lasinio model with AMM}\label{formalism}

The two-flavor NJL model in an external electromagnetic field and constant anomalous magnetic moment is given by the following Lagrangian \cite{Fayazbakhsh:2014mca}

\begin{eqnarray}
\mathcal{L}&=&\overline{\psi}\left(i \slashed D - \hat{m}+\frac{1}{2}\hat{a}\sigma^{\mu\nu}F^{\mu\nu}\right)\psi\nonumber\\
           &&+G\left[(\overline{\psi}\psi)^{2}+(\overline{\psi}i\gamma_{5}\vec{\tau}\psi)^{2}\right],\label{lagrangian}
\end{eqnarray}

\noindent where $G$ is the coupling 
constant, $A^\mu$ the electromagnetic gauge field, $F^{\mu\nu} = \partial^\mu A^\nu - \partial^\nu A^\mu$ is the electromagnetic tensor,  $\vec{\tau}$ are isospin Pauli matrices,  $Q$ is the diagonal quark charge matrix\footnote{Our results are expressed in Gaussian natural units 
where $1\,{\rm GeV}^2= 1.44 \times 10^{19} \, G$ and $e=1/\sqrt{137}$.} 
[$Q= \ $diag($q_u = 2 /3$, $q_d = -1/3$)],
$D^\mu =(\partial^{\mu} + i e Q A^{\mu})$ is the covariant derivative, the bare quark mass matrix is $\hat{m}=\ $diag($m_u$, $m_d$), and  
 $\psi=(\psi_u \quad \psi_d)^T$ is the quark fermion field. 
We consider the isospin limit $m_u = m_d = m$  and the Landau gauge,  $A_{\mu}=\delta_{\mu 2}x_{1}B$, in order to satisfy 
 $\nabla \times \vec{A}=\vec{B}=B{\hat{e_{3}}}$ and $\nabla \cdot \vec{A}=0$. 

The Lagrangian $\mathcal{L}$ in the mean field approximation is given by

\begin{equation}
 \mathcal{L}=\overline{\psi}\left(i\slashed{D}-M+\frac{1}{2}\hat{a}\sigma_{\mu\nu}F^{\mu\nu}\right)\psi - \frac{(M-m)^2}{4G},
\end{equation}

\noindent where we have redefined the mass term, now as a constituent quark mass 
\begin{equation}
 M=m-2G \left \langle \overline{\psi}\psi \right \rangle. \label{gap}
\end{equation}

In the last equation, $\left \langle \overline{\psi}\psi \right \rangle$ is the chiral condensate.
The AMM appears adding the phenomenological Pauli term, $\frac{1}{2}\hat{a}\sigma^{\mu\nu}F^{\mu\nu}$ term, which couples the AMM of the quarks, given through the entries of the matrix $\hat{a}$ for each flavor $f=u,d$ as $\hat{a}= \ $diag($a_u$, $a_d$), with the external magnetic field.  
The previous quantities are given, in the one-loop level approximation, by

\begin{eqnarray}
a_f= q_f \alpha_f \mu_B, \; \alpha_f=\frac{\alpha_e q_f^2}{2\pi}, \;  \alpha_e=\frac{1}{137},\; \mu_B=\frac{e}{2M} \;,
\end{eqnarray}

\noindent where $\alpha_e$ is the fine-structure constant and $\mu_B$ is the Bohr magneton. As we will work with constant values of AMM we can redefine $\mu_B=e/2M \rightarrow \mu_B =e/2M_0$, where $M_0 = M(T=0,eB=0)$~\cite{Fayazbakhsh:2014mca}.

\section{Thermodynamic potential with quark AMM}\label{pot_eff}

The thermodynamic potential of the two-flavor NJL model at zero temperature with the influence of quark AMM is given by \cite{Fayazbakhsh:2014mca}
\begin{eqnarray}
 \Omega = \frac{(M-m)^2}{4G}+\Omega^{\text{mag}}(M,B),
\end{eqnarray}

\noindent where $\Omega^{\text{mag}}$ is the magnetic contribution given by

{\small
\begin{eqnarray}
 \Omega^{\text{mag}}(M,B) = -N_c\sum_{f=u,d}|B_f|\sum_{n=0}^{\infty}\sum_{s=\pm 1}\int_{-\infty}^{\infty}\frac{dp_3}{4\pi^2}E_{n,s}^f \;,\label{omega1}
\end{eqnarray} }

\noindent in which the  $B_f \equiv q_feB$, $s=\pm 1$ is the spin index, $n$ are the Landau levels, $N_c = 3$ the number of colors and the quark-energy dispersion relation is defined as

\begin{eqnarray}
 E_{n,s}^f=\sqrt{p_3^2+(M_{n,s}^f-sa_fB)^2},
\end{eqnarray}

\noindent where $M_{n,s}^f=\sqrt{|B_f|(2n+1-s_fs)+M^2}$ and $s_f=\text{sign}(q_f)$ is the charge sign function. Using the gamma function integral representation, one may write: 

\begin{eqnarray}
 \frac{1}{A^n}=\frac{1}{\Gamma(n)}\int_0^{\infty}d\tau\tau^{n-1}e^{-\tau A} \; .
\end{eqnarray}

Thus, after some straightforward manipulations, we can rewrite the magnetic part of the thermodynamic potential, Eq. (\ref{omega1}) as

\begin{eqnarray}
 \Omega^{\text{mag}}(M,B) &=& \frac{N_c}{8\pi^2}\sum_{f=u,d}|B_f|\nonumber \\
 &\times&\sum_{n=0}^{\infty}\sum_{s=\pm 1}\int_0^{\infty}\frac{d\tau}{\tau^2}e^{-\tau(M_{n,s}^f-sa_fB)^2} \nonumber\\
 &=& \frac{N_c}{8\pi^2}\sum_{f=u,d}\int_0^{\infty}\frac{d\tau}{\tau^3}e^{-\tau M^2}F_f(\tau), \label{omega2}
\end{eqnarray}

\noindent where we have defined

\begin{eqnarray}
 F_f(\tau)&=&e^{-\tau(a_fB)^2}\tau |B_f|\nonumber\\ 
 &\times&\sum_{n=0}\sum_{s=\pm 1}e^{-\tau\left(|B_f|(2n+1-s_fs)-2sa_fBM_{n,s}^f\right)}.\label{Ftau}\quad
\end{eqnarray}

It is possible to rewrite the function $F_f(\tau)$, as detailed in Appendix A, in the following way

\begin{eqnarray}
F_f(\tau) &=& e^{-\tau(a_fB)^2}\tau|B_f|\left[s_f\sinh(\tau2a_fBM) \right. \nonumber\\
&& \left. +\sum_{k=0}^{\infty}\frac{(2a_fB)^{2k}}{(2k)!}(-1)^k\tau^{2k}D_k(\tau)\right]\nonumber\\
 &=& e^{-\tau(a_fB)^2}\tau|B_f|\left[s_f\sinh(\tau2a_fBM) \right. \nonumber\\
&& \left. + F^{(2)}_f(\tau)\right]\label{Dk0}.
\end{eqnarray}

The function $F^{(2)}(\tau)$ is 

\begin{eqnarray}
    F^{(2)}_f(\tau) = \sum_{k=0}^{\infty}\frac{(2a_fB)^{2k}}{(2k)!}(-1)^k\tau^{2k}D_k(\tau),\label{F2}
\end{eqnarray}

\noindent and will be useful in the next sections. The function $D_k(\tau)$ is given by

\begin{eqnarray}
D_k(\tau)&=&e^{\tau M^2}\frac{d^k}{d\tau^k}e^{-\tau M^2}\coth(\tau|B_f|)\nonumber\\
&=& \sum_{n=0}^{k} \binom{k}{n}(-M^2)^{k-n}\frac{d^n}{d\tau^n}\coth(|B_f|\tau)\nonumber\\
&=&(-1)^k(M^2)^k\sum_{n=0}^{k} \binom{k}{n}(-1)^{n}\left(\frac{|B_f|}{M^2}\right)^{n}\nonumber \\ 
&&\times\frac{d^n}{d(\tau |B_f|)^n}\coth(|B_f|\tau).\label{Dk1}
\end{eqnarray}

This last representation is useful to express the function $D_k(\tau)$ as a series expansion in terms of the parameter $x_f=|B_f|/M^2$.

\section{Mass-dependent regularization}
\label{MDreg}

In order to apply the VMR scheme \cite{Avancini:2020xqe}, we have to identify the divergences of the thermodynamic potential, Eq. (\ref{omega1}).
It is clear that the integral diverges in the limit of $\tau \to 0$.
It is possible to understand the origin of the divergence considering the Taylor expansion of the function $F_f(\tau)$ around $\tau=0$ up to the order $\mathcal{O}(\tau^2)$ as

\begin{eqnarray}
 F^0_f(\tau)&=&1+(a_fB)^2\tau+R_f(B_f,M)\tau^2\nonumber\\ 
 && +\mathcal{O}(\tau^3),\; \tau\ll1, \label{Ftau0}
\end{eqnarray}

\noindent where the coefficient of $\tau^2$ is mass-dependent and given by

\begin{eqnarray}
 R_f(B_f,M)&=&\frac{|B_f|^2}{3}-\frac{(a_fB)^4}{6}+2(a_fB)^2M^2\nonumber\\
 &&+s_f2|B_f|(a_fB)M.
\end{eqnarray}
To regularize the effective potential, we will apply the VMR prescription \cite{Avancini:2020xqe}:

\begin{eqnarray}
\Omega^{\text{mag}}(B,M) &=& [\Omega^{\text{mag}}(B,M) - \Omega^{VD}(B,M)] \nonumber\\
&+& \Omega^{VD}(B,M)\nonumber\\
 &\rightarrow& \Omega^{\text{mag}}_R(B,M) + \Omega^{VM}(B,M),
\end{eqnarray}

\noindent where $\Omega^{\text{mag}}_R = \Omega^{\text{mag}}(B,M) - \Omega^{VD}(B,M)$ is the magnetic part of the regularized thermodynamic potential, $\Omega^{VD}(B,M)$ is the vacuum-divergent contribution and $\Omega^{VM}(B,M)$ is the vacuum-magnetic contribution. The last two quantities are given by

\begin{eqnarray}
 \Omega^{VD}(B,M) = \frac{N_c}{8\pi^2}\sum_{f=u,d}\int_0^{\infty} \frac{d\tau}{\tau^3}e^{-\tau M^2}F_f^0(\tau),\nonumber \\
 \Omega^{VM}(B,M) = \frac{N_c}{8\pi^2}\sum_{f=u,d}\int_{\frac{1}{\Lambda^2}}^{\infty} \frac{d\tau}{\tau^3}e^{-\tau M^2}F_f^0(\tau).\label{omegaVM}
\end{eqnarray}

In the presence of the AMM, the vacuum-magnetic term now has mass-dependent contributions which makes its behavior completely different from the usual case where we have zero AMM.  In that case, we have simply

\begin{eqnarray}
 F_f(\tau)&=&|B_f|\tau\coth(|B_f|\tau)\nonumber\\
        &\sim&  1+\frac{(|B_f|\tau)^2}{3}, \quad \tau\ll 1.
\end{eqnarray}

 In the latter case, the full regularization will include new MI terms proportional to $eB^2$ and these new contributions do not give additional physics \cite{Avancini:2020xqe}. Now, in the nonzero AMM case, if we look at Eq. (\ref{omega2}), the thermodynamic potential has divergences until $\mathcal{O}(\tau^2)$ and, consequently, the gap equation until $\mathcal{O}(\tau)$. 
In this way, for the gap equation, we just need to use the expansion given in Eq. (\ref{Ftau0}) including first-order terms in $\tau$, $F_f^0(\tau)=1+(a_fB)^2\tau$, which seems not to be consistent with the constraint, $\partial \Omega/\partial M=0$, when we start from the regularized thermodynamic potential, including the additional mass-dependent contribution of order $\mathcal{O}(\tau^2)$. Therefore, these additional mass-dependent contributions are enforcing the regularization to remove unnecessary new terms in the gap equation.

\section{Numerical Results}\label{results}
In this work we make use of the following parametrization of the two-flavor NJL model using the proper-time vacuum regularization: $\Lambda = 886.62$ MeV, $m_0=7.383$ MeV and $G = 4.001/\Lambda^2$, in order to reproduce the quark condensate, $\langle \bar{u}u\rangle^{1/3} = -220$ MeV, the pion mass, $m_{\pi}=138$ MeV and the pion decay constant, $f_{\pi}=92.4$ MeV \cite{Avancini:2019wed}.

\begin{figure}[ht]
\begin{tabular}{ccc}
\includegraphics[width=9.0cm]{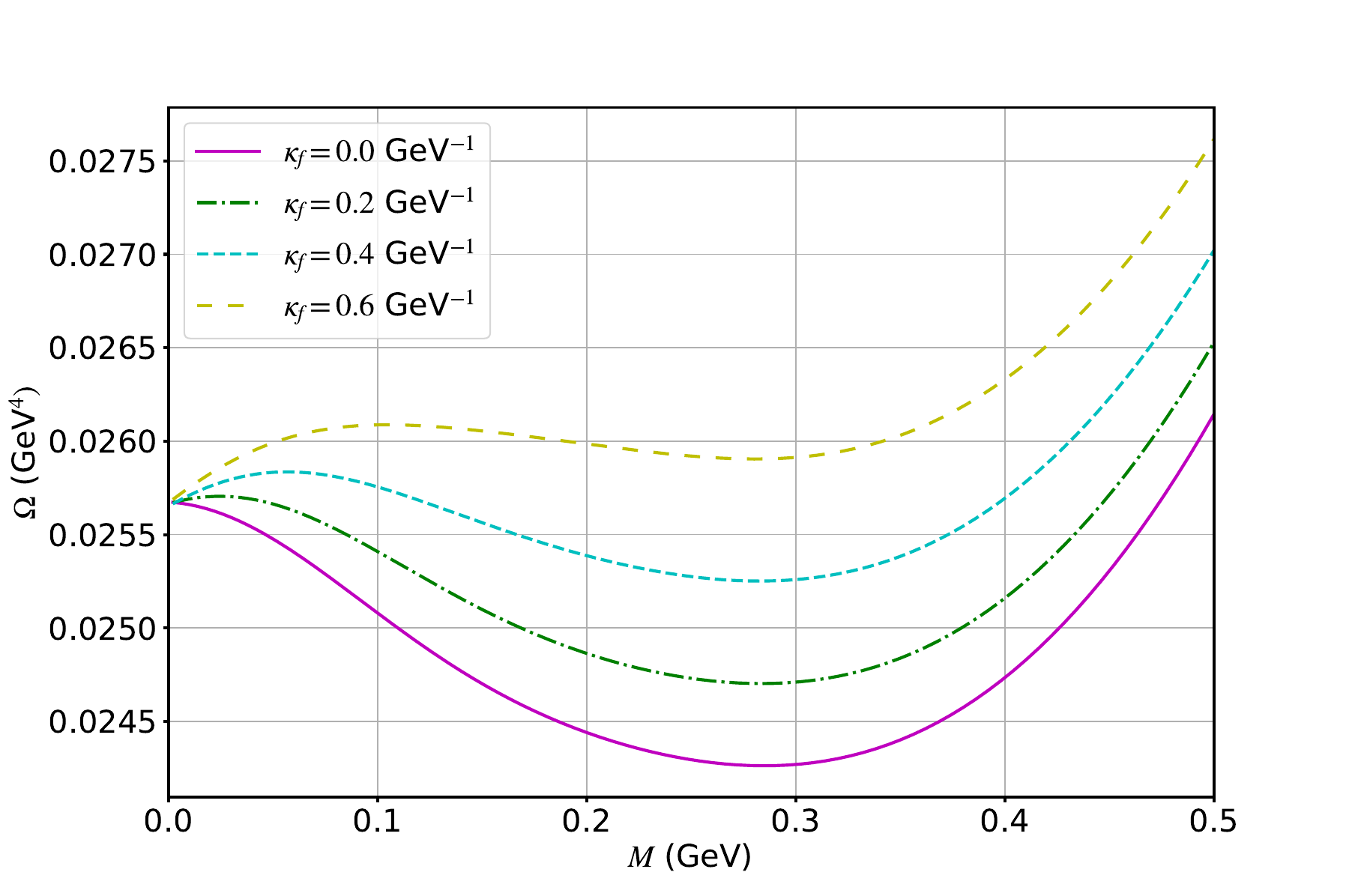}\\
\end{tabular}
\caption{Thermodynamic potential as a function of the effective quark mass for different values of quark AMM with $2(a_fB)^2M^2+s_f2Ba_fM\neq 0$ at $eB=0.3$ GeV$^2$.}
\label{potMD}
\end{figure}

In Fig. \ref{potMD} we show the thermodynamic potential including mass-dependent terms in the regularization as a function of the effective quark mass for different values of $\kappa_f$ with a fixed magnetic field value, $eB=0.3 \text{ GeV}^2$. For lower values of $\kappa_f$, we have a global minimum at $M\sim 0.3$ GeV, however, as we increase the value of $\kappa_f$, the minimum at $M=0$ surpasses the massive one, and we have the first-order phase transition observed before in the literature. This behavior also occurs for some fixed value of $\kappa_f$ and strong magnetic field values, as observed in Ref. \cite{Fayazbakhsh:2014mca}.

\begin{figure}[ht]
\begin{tabular}{ccc}
\includegraphics[width=9.0cm]{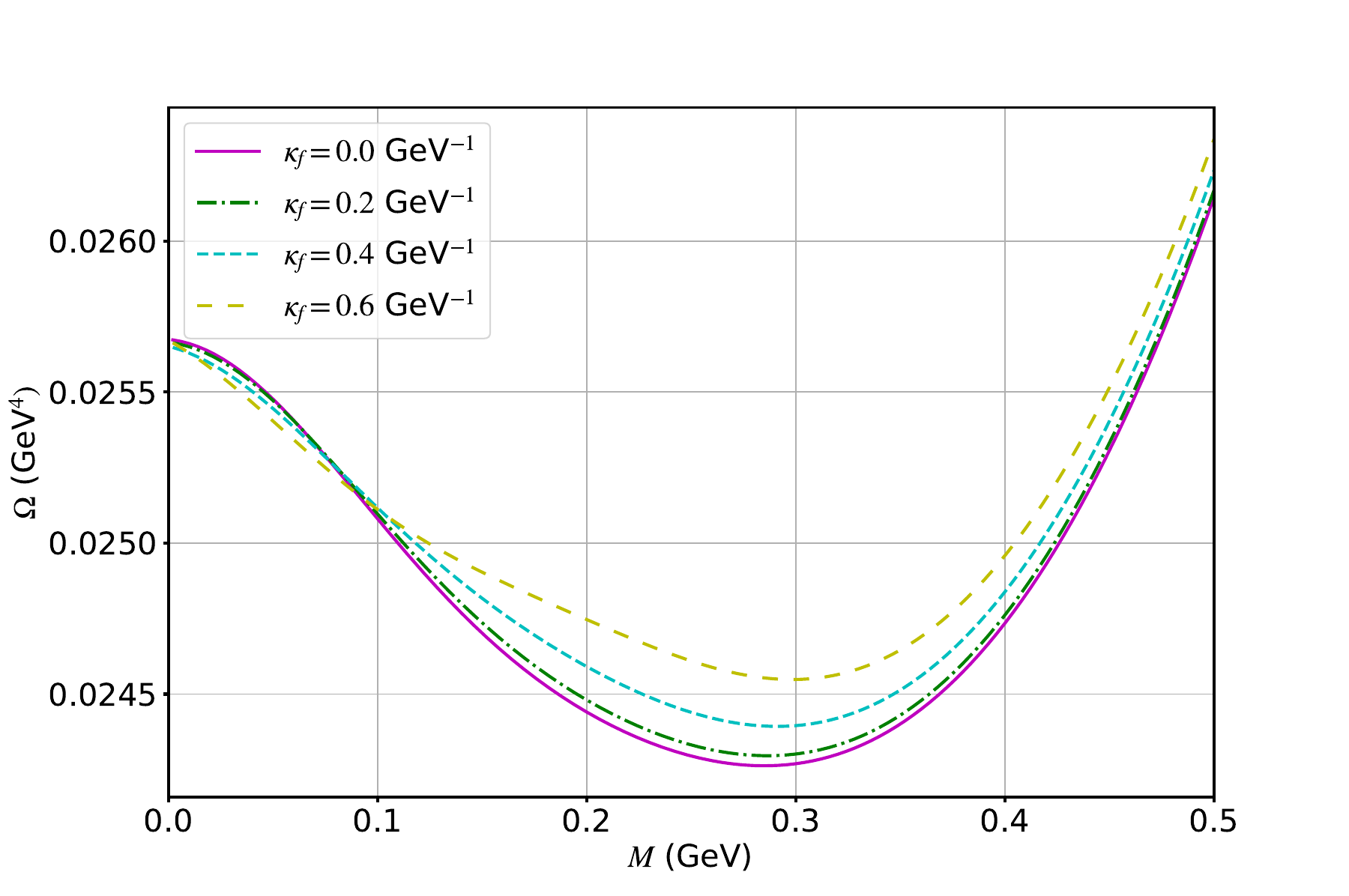}\\
\end{tabular}
\caption{Thermodynamic potential as a function of the effective quark mass for different values of quark AMM with $2(a_fB)^2M^2+s_f2Ba_fM=0$ at $eB=0.3$ GeV$^2$.}
\label{potMI}
\end{figure}

The same quantities are plotted in Fig. \ref{potMI}, in which we test $\Omega^{VM}$ assuming $2(a_fB)^2M^2+s_f2Ba_fM=0$. We see that the usual first-order phase transition does not occur when we increase the value of $\kappa_f$, showing that, these terms are the ones mainly responsible for unbalancing the thermodynamic potential. 


\section{Mass-independent regularization}\label{secMI}

In order to show a way to avoid nonphysical behavior of the model due to mass-dependent terms in the regularization, we propose in this section the mass-independent regularization, already explored in detail in Ref. \cite{Farias:2021fci}.
The Schwinger {\it ansatz} proposed in Eq. (\ref{lagrangian}) for AMM is applied to the one-loop correction for QED in the limit of $x_f\rightarrow 0$.  The same idea must be used here, so we will apply the $\mathcal{O}(x_f^0)$ expansion in the thermodynamic potential, Eq. (\ref{omega2}). To see this, we rewrite the summation in Eq. (\ref{F2}) as

\begin{eqnarray}
F_f^{(2)} &=& \sum_{k=0}^{\infty}\sum_{n=0}^k\binom{k}{n}\frac{(2a_fB\tau M)^{2k}}{(2k)!}(-1)^n\left(\frac{|B_f|}{M^2}\right)^{n}\overline{D_n}(\tau),\nonumber 
\end{eqnarray}

\noindent where $\overline{D_n}(\tau) = \frac{d^n}{d(\tau|B_f|)^n}\coth(|B_f|\tau)$. Since, phenomenologically we implement $a_f = q_fe\kappa_f$ and $\kappa_f=\alpha_f/2M_0$, where $M_0\equiv M(B=0,T=0)$, we can identify

\begin{widetext}
\begin{eqnarray}
F_f^{(2)} &=& \sum_{n,k}\binom{k}{n}\frac{(\alpha_fB_f\tau )^{2k}}{(2k)!}\left(\frac{M}{M_0}\right)^{2k}(-1)^n\left(\frac{|B_f|}{M^2}\right)^{n}\overline{D_n}(\tau)\nonumber\\
&=& \sum_{n,k}\binom{k}{n}\frac{(\alpha_fB_f\tau )^{2k}}{(2k)!}\left(\frac{M_0 + \delta M}{M_0}\right)^{2k}(-1)^n\left(\frac{|B_f|}{(M_0 + \delta M)^2}\right)^{n}\overline{D_n}(\tau)\nonumber\\
&=& \sum_{n,k}\binom{k}{n}\frac{(\alpha_fB_f\tau )^{2k}}{(2k)!}\left(1+\epsilon\right)^{2k} (-1)^n\left(\frac{|B_f|}{M_0^2(1 + \epsilon)^2}\right)^{n}\overline{D_n}(\tau)\nonumber\\
&\approx& \sum_{n,k}\binom{k}{n}\frac{(\alpha_fB_f\tau )^{2k}}{(2k)!}\left(1+2k\epsilon\right)(1 - 2n\epsilon) (-1)^n\left(\frac{|B_f|}{M_0^2}\right)^{n}\overline{D_n}(\tau)\nonumber\\
&=& \sum_{n,k}\binom{k}{n}\frac{(\alpha_fB_f\tau )^{2k}}{(2k)!}(1 + 2(k-n)\epsilon) (-1)^n\left(\frac{|B_f|}{M_0^2}\right)^{n}\overline{D_n}(\tau),
\end{eqnarray}

\noindent where $\epsilon =\delta M/M_0$ represents how much $M$ changes in relation to $M_0$. At this point, we assume $|B_f|/M_0\ll1$, so the $n=0$ term is dominant. In this way, we have

\begin{eqnarray}
 F_f^{(2)} &\approx&\sum_{k}\frac{(\alpha_fB_f\tau )^{2k}}{(2k)!}(1 + 2k\epsilon)\coth(|B_f|\tau)\nonumber\\
 &=&\sum_{k}\frac{(\alpha_fB_f\tau )^{2k}}{(2k)!}\coth(|B_f|\tau)+\sum_{k}\frac{(\alpha_fB_f\tau )^{2k}}{(2k)!}( 2k\epsilon)\coth(|B_f|\tau)\nonumber\\
 &=&\coth(|B_f|\tau)[\cosh(\alpha_f|B_f|\tau)+\epsilon\tau\frac{\partial}{\partial \tau}\cosh(\alpha_f|B_f|\tau)]\nonumber\\
 &=&\coth(|B_f|\tau)[\cosh(\alpha_f|B_f|\tau)+\epsilon\tau\alpha_f |B_f|\sinh(\alpha_f|B_f|\tau)].\nonumber\\
\end{eqnarray}

\end{widetext}
It is easy to show that the function $F_f(\tau)$ is now given by

\begin{eqnarray}
 F_f(\tau)&=& e^{-\tau(a_fB)^2}\tau|B_f|\left[s_f\sinh(\tau2a_fBM) \right.\nonumber\\
 &&+\left. \cosh(\tau2a_fBM)\coth(|B_f|\tau)\right]\nonumber\\
 &&+\coth(|B_f|\tau)e^{-\tau(a_fB)^2}\tau|B_f|\nonumber\\ &&\times\left[\epsilon\tau\alpha_f |B_f|\sinh(\alpha_f|B_f|\tau)\right]\label{Ftau1}.
\end{eqnarray}

The term proportional to $\epsilon$ can be integrated in Eq. (\ref{omega2}) as

\begin{eqnarray}
 &&\epsilon\alpha_f|B_f|^2\int_{\varepsilon}^{\infty} \frac{d\tau}{\tau^3}e^{-\tau M^2}e^{-\tau(a_fB)^2}\tau^2\sinh(\alpha_f|B_f|\tau)\nonumber\\
 &&\times\coth(|B_f|\tau) \rightarrow\epsilon\alpha_f|B_f|^2\int_{\varepsilon'}^{\infty} \frac{d\tau}{\tau}e^{-\tau}e^{-\tau\frac{(a_fB)^2}{M^2}}\nonumber\\
&&\times\sinh\left(\frac{\alpha_f|B_f|}{M^2}\tau\right)\coth\left(\frac{|B_f|\tau}{M^2}\right)\nonumber,
\end{eqnarray}

\noindent where we have done a simple change of variables. To simplify the analysis, we assume that we have regulated the region $\tau\sim0$ with some generic regulator $\varepsilon$. We can normalize the effective potential by $M_0^4$ in order to obtain

{\small\begin{eqnarray}
\lambda_f\int \frac{d\tau}{\tau}e^{-\tau}e^{-\tau\frac{(a_fB)^2}{M^2}}\sinh\left(\frac{\alpha_f|B_f|}{M^2}\tau\right)\coth\left(\frac{|B_f|\tau}{M^2}\right) \rightarrow0,\nonumber\\ \label{integral}
\end{eqnarray}}

\noindent in the limit which $\lambda_f=\epsilon\alpha_f\left(\frac{|B_f|}{M_0^2}\right)^2\ll1$. The last expansion is a bonus result, which is useful for physical situations where $M$ can increase its value, as at low temperatures and finite magnetic fields, and situations where $M$ decreases, as at finite temperatures. It is easy to show that, for usual parameters of the NJL model, such as the one listed in the numerical results section, as well as the values for the quark AMM \cite{Fayazbakhsh:2014mca}, the previous integration is very small when compared to the mean field term of the thermodynamical potential. With this analysis, we guarantee the convergence of two situations:
\begin{enumerate}[label=(\roman*)]
 \item $\delta M > 0$: it can happen at finite $eB$ and low temperatures. In the extremal situation, where $T=0$, we can simplify the analysis using the expansion at $B_f/M^2<1$.
 
 \item $\delta M < 0$: it can happen at finite $eB$ and high values of temperature. The expansion parameter $\epsilon\lesssim 1$, depending the value of $\alpha_f$. However, we still guarantee the values of $(|B_f/M_0^2|)\alpha_f\epsilon\ll 1$ .
 \end{enumerate}

After some straightforward steps

\begin{eqnarray}
 F_f(\tau)&=& e^{-\tau(a_fB)^2}\tau|B_f|\left[\frac{\cosh((\alpha_f+1)|B_f|\tau)}{\sinh(|B_f|\tau|)}\right]\nonumber\label{Ftau2}.
\end{eqnarray}

The magnetic field part of thermodynamic potential is then given by

\begin{eqnarray}
 \Omega^{\text{mag}}= \frac{N_c}{8\pi^2}\sum_f\int_0^{\infty}\frac{d\tau}{\tau^3}e^{-\tau \mathcal{K}_{0,f}^2}\left[\tau|B_f|\frac{\cosh(c_f|B_f|\tau)}{\sinh(|B_f|\tau|)}\right],\nonumber\\
\end{eqnarray}

\noindent where we have defined $c_f = a_f+1$ and $\mathcal{K}_{0,f}=\sqrt{M^2+(a_fB)^2}$. This last expression is exactly the same used in \cite{Farias:2021fci} adapting to the NJL model the one-loop Schwinger-Weisskopf Lagrangian in \cite{Dittrich:1977ee}. In order to regularize this latter thermodynamic potential, we expand the hyperbolic functions at $\tau\sim 0$

\begin{eqnarray*}
 \frac{\cosh(c_fB_f\tau)}{\sinh(B_f\tau)}\sim\frac{1}{B_f\tau}+\frac{B_f\tau}{6}\left(3c_f^2-1\right)+\mathcal{O}(\tau^3).
\end{eqnarray*}

The expansion of $F_f^0(\tau)$ is then, given by

\begin{eqnarray}
 F_f^0(\tau)=1+\frac{(B_f\tau)^2}{6}\left(3c_f^2-1\right)+\mathcal{O}(\tau^3)
\end{eqnarray}

\noindent which is mass independent. In this case, the gap equation have the correct degree of divergence as the zero AMM case, avoiding, therefore, the contradiction observed in the mass-dependent regularization.

\section{Conclusions}\label{conclusions}

In this work, we have shown the effects of the quark AMM in the context of the two-flavor Nambu--Jona-Lasinio model in a strong magnetic field. The inclusion of the quark AMM is given through a set of constant values, by using the Schwinger {\it ansatz} in the NJL Lagrangian. We then explore the role of the possible first-order phase transition in the thermodynamic potential that can occur when the value of quark AMM is strong enough. To understand the origin of this transition, we apply the VMR, since the subtraction of divergences shows clearly the behavior of the model in the ultraviolet limit. We identify that the full VMR prescription regularizes MD terms that uneven the thermodynamic potential at $M=0$, which is strengthened for higher values of quark AMM until an apparent first-order phase transition takes place. We argue that these MD terms are nonphysical and should be avoided by using a MI regularization. In the limit of $|B_f|/M^2\ll 1$, it is shown that we can obtain the usual Schwinger-Weisskopf approach for one-loop approximation in QED applied to the NJL model. In this approach, the MI terms are obtained in the regularization of the vacuum-magnetic contributions and the results indicate the absence of first-order phase transitions in accordance
with predictions of lattice QCD \cite{Bali:2012zg}.

\section*{ACKNOWLEDGMENTS}

This work was partially supported by Conselho Nacional de Desenvolvimento Cient\'ifico 
e Tecno\-l\'o\-gico  (CNPq), Grants No. 309598/2020-6 (R.L.S.F.), No. 304518/2019-0 (S.S.A.); Coordena\c c\~{a}o  de 
Aperfei\c coamento de Pessoal de  N\'{\i}vel Superior - (CAPES) Finance  Code  001 and Fundação Carlos Chagas Filho de Amparo à Pesquisa do 
Estado do Rio de Janeiro (FAPERJ), Grant No.SEI-260003/019544/2022 (W.R.T); 
Funda\c{c}\~ao de Amparo \`a Pesquisa do Estado do Rio 
Grande do Sul (FAPERGS), Grants Nos. 19/2551- 0000690-0 and 19/2551-0001948-3 (R.L.S.F.);  
CAPES, Grant no. 88887.826087/2023-00 (R.P.C); 
The work is also part of the project Instituto Nacional de Ci\^encia 
e Tecnologia - F\'isica Nuclear e Aplica\c{c}\~oes (INCT - FNA), Grant No. 464898/2014-5. 

\appendix
\section{Calculating $\boldsymbol{F_f(\tau)}$}\label{appA}
In this section, we will obtain the expression to the $F_f(\tau)$, Eq. (\ref{Dk0}). 
Separating term proportional to $a_f$ in the exponential of the Eq. (\ref{Ftau})
and writing it in a Taylor expansion,

\begin{eqnarray}
F_f(\tau) 
&=& e^{-\tau(a_f B)^2} \tau |B_f| \sum_{k=0}^{\infty} \frac{(\tau 2 a_f B)^k}{k!}  \nonumber\\
&\times&
\sum_{n=0}^{\infty}\sum_{s=\pm 1} s^k e^{-\tau |B_f|(2n+1-s_f s)} 
\left (M_{n,s}^f\right)^{k}.\nonumber\\
\label{F_f}
\end{eqnarray}

We can explicitly do the summation on even and odd parts in respect to $k$, like

\begin{eqnarray}
&&F_f(\tau) 
= e^{-\tau(a_f B)^2} \tau |B_f| 
 \nonumber\\
&\times&\sum_{k=0}^{\infty} \left \{ \frac{(\tau 2 a_f B)^{2k}}{2k!}\sum_{n=0}^{\infty}\sum_{s=\pm 1} 
e^{-\tau |B_f|(2n+1-s_f s)} \left (M_{n,s}^f\right)^{2k}  \right. \nonumber \\
&+&\left . \frac{( \tau  2 a_f B)^{2k+1}}{(2k+1)!}  
\sum_{n=0}^{\infty}\sum_{s=\pm 1}s \ 
e^{-\tau |B_f|(2n+1-s_f s)} \left (M_{n,s}^f\right)^{2k+1}
\right\},\nonumber\\
\label{F_f_even_odd}
\end{eqnarray}
where the even power of $s$ is given by $s^{2k}=1$ and the odd $s^{2k+1} = s$. Analyzing the
summations over $n$ and $s$ to the odd part of last expression in the case where $s_f = 1$,
\begin{widetext}
\begin{eqnarray}
\sum_{n=0}^{\infty}\sum_{s=\pm 1} 
s \ e^{-\tau |B_f|(2n+1-s)} [|B_f|(2n+1-s)+M^2]^{\frac{2k+1}{2}}
&=&\underbrace{\left[M^2 \right ]^{\frac{2k+1}{2}}}_{n=0, s=1}-
\underbrace{e^{-\tau |B_f|2} \left [|B_f|2+M^2 \right ]^{{\frac{2k+1}{2}}}}_{n=0, s=-1}\nonumber\\
&+&\underbrace{e^{-\tau |B_f|2} \left [|B_f|2+M^2 \right ]^{{\frac{2k+1}{2}}}}_{n=1, s=1}
-\underbrace{e^{-\tau |B_f|4} \left [|B_f|4+M^2 \right ]^{{\frac{2k+1}{2}}}}_{n=1, s=-1} \nonumber\\
&+&\underbrace{e^{-\tau |B_f|4} \left [|B_f|4+M^2 \right ]^{{\frac{2k+1}{2}}}}_{n=2, s=1}
-\underbrace{e^{-\tau |B_f|6} \left [|B_f|6+M^2 \right ]^{{\frac{2k+1}{2}}}}_{n=2, s=-1}+\text{...} \nonumber \\
&=& M^{2k+1}+\sum_{s=\pm 1} s\sum_{n=1}^{\infty} e^{-\tau|B_f|2n} [|B_f| 2n+ M^2]^{{\frac{2k+1}{2}}}\nonumber\\
&=& M^{2k+1},
\label{LHS}
\end{eqnarray}
\end{widetext}
where next to the last line in Eq. (\ref{LHS}) the summation goes to zero because of the
explicitly $s$. In the case where $s_f = -1$ is easy to see that the result is analogous and given
by $-M^{2k+1}$. So, all the summation of the odd part of Eq. (\ref{F_f_even_odd}) has
$s_f M^{2k+1}$ as the result. To the even summation, we insert an identity 
$e^{\tau M^2 - \tau M^2}$, then
\begin{eqnarray*}
&& 
\sum_{n=0}^{\infty}\sum_{s=\pm 1} 
e^{-\tau |B_f|(2n+1-s_f s)} [|B_f|(2n+1-s)+M^2]^{k} \nonumber\\
&& =e^{\tau M^2} (-1)^k \frac{d^k}{d\tau^k} \left\{ e^{-\tau M^2} 
\sum_{n=0}^{\infty}\sum_{s=\pm 1} e^{-\tau \left[|B_f|(2n+1-s_f s)\right]}  
 \right\}.
\end{eqnarray*}
From the Ref. \cite{Bubnov:2017}, we can show that
\begin{eqnarray*}
    \sum_{n=0}^{\infty}\sum_{s=\pm 1} e^{-\tau \left[|B_f|(2n+1-s_f s)\right]}  &=&
    1 + 2 \sum_{n=1}^{\infty} e^{-\tau |B_f| 2n} \nonumber \\
    &=& \coth{(\tau |B_f|)}.
\end{eqnarray*}
Plugging all the contribution we have the explicit summation of the $\sinh(\tau 2 a_f B M)$ and so
Eq. (\ref{F_f}) becomes
\begin{eqnarray}
F_f(\tau) &=& 
e^{-\tau(a_f B)^2} \tau |B_f|\left\{ 
\vphantom{\frac{1}{1}}
s_f \sinh(\tau 2 a_f B M) \right . \nonumber \\
&+&\left . \sum_{k=0}^{\infty} \frac{(\tau 2 a_f B)^{2k}}{(2k)!} 
(-1)^k e^{\tau M^2}\right.\nonumber\\
&\times&\left.\frac{d^k}{d\tau^k} e^{-\tau M^2} \coth(\tau|B_f|)\right \}. \nonumber \\
\end{eqnarray}

\bibliographystyle{spphys}
\bibliography{AMM.bib}

\end{document}